\documentclass{aa}

\usepackage{graphicx}
\usepackage{txfonts}
\usepackage[inline]{enumitem}
\usepackage{multirow}
\usepackage[colorlinks=true, linkcolor=blue, citecolor=blue, filecolor=blue, urlcolor=blue]{hyperref}

\begin{document}

   \title{The effect of local and large scale environment on the star formation histories of galaxies}
   \titlerunning{The effect of local and large scale environment on the star formation histories of galaxies}


    \author{G. Torres-Ríos \inst{1} \and 
           I. Pérez \inst{1, 2} \and  
           S. Verley \inst{1, 2} \and
           J. Domínguez-Gómez \inst{1} \and
           M. Argudo-Fernández \inst{1, 2} \and
           S. Duarte Puertas \inst{1, 2, 3} \and
           A. Jiménez \inst{1} \and
           T. Ruiz-Lara \inst{1, 2} \and
           A. Zurita \inst{1, 2} \and
           B. Bidaran \inst{1} \and
           A. Conrado \inst{4} \and
           D. Espada \inst{1,2} \and
           R. García-Benito \inst{4} \and
           R. M. González Delgado \inst{4} \and
           J. Falcón-Barroso \inst{5,6} \and
           E. Florido \inst{1,2} \and
           P. Sánchez-Blázquez \inst{7} \and
           L. Sánchez-Menguiano \inst{1, 2}           
          }

   \institute{
   Universidad de Granada, Departamento de Física Teórica y del Cosmos, Campus Fuentenueva, Edificio Mecenas, 18071 Granada, Spain.
   \email{gloriatr@ugr.es}
      \and
      Instituto Carlos I de F\'isica Te\'orica y Computacional, Facultad de Ciencias, 18071 Granada, Spain
      \and
      Département de Physique, de Génie Physique et d’Optique, Université Laval, and Centre de Recherche en Astrophysique du Québec (CRAQ), Québec, QC, G1V 0A6, Canada
      \and
      Instituto de Astrofísica de Andalucía - CSIC, Glorieta de la Astronomía s.n., 18008 Granada, Spain
      \and
      Instituto de Astrofísica de Canarias, Vía Láctea s/n, 38205 La Laguna, Tenerife, Spain
      \and
      Departamento de Astrofísica, Universidad de La Laguna, 38200 La Laguna, Tenerife, Spain
      \and
      Departamento de Física de la Tierra y Astrofísica IPARCOS, UCM, 28040 Madrid, Spain
             }

   \date{Received May 10, 2024; accepted September 27, 2024}
  
  \abstract
  {The environment of galaxies may play a key role in their evolution. Large extragalactic surveys makes it possible to study galaxies not only within their local environment, but also within the large-scale structure of the Universe.} 
   {We aim to investigate how the local environment influences the star formation history (SFH) of galaxies residing in various large-scale environments.}
   {We categorise a sample of 9384 galaxies into the three primary large scale structures (voids, walls \& filaments, and clusters) and further classify them based on their local environment (as either "singlets" or group members), through a search of companion galaxies within sky-projected distances $\Delta r_p < 0.45$ Mpc and velocity differences $\Delta v < 160$ $\text{km s}^{-1}$. Subsequently, we explore these subsamples through SFH data from previous works. Throughout the study, galaxies are divided into long-timescale SFH galaxies (LT-SFH), which assemble their mass steadily along cosmic time, and short-timescale SFH galaxies (ST-SFH), which form their stars early. We then compare characteristic mass assembly look-back times.}
  {The distributions of mass assembly look-back times in ST-SFH galaxies are statistically different for singlets and groups. These differences are only found in LT-SFH galaxies when studying these distributions in stellar mass bins. Our results indicate that the large-scale environment is related to a delay in mass assembly of up to $\sim$2 Gyr, while this delay is $<$1 Gyr in the case of local environment. The effect of both kinds of environment is more significant in less massive galaxies, and in LT-SFHs.}
   {Our results are consistent with galaxies in groups assembling their stellar mass earlier than singlets, especially in voids and lower mass galaxies. Local environment plays a relevant role in stellar mass assembly times, although we find that large-scale structures also cause a delay in mass assembly, more so in the case of cluster galaxies.}


   \maketitle
%

\section{Introduction}

    Mass assembly and star formation in galaxies constitute a pivotal focus within the study of galaxy evolution. Environment plays a relevant role in these processes, and it is now, in the time of large extragalactic surveys, possible to examine them thoroughly across different spatial, extragalactic scales. Within the context of the large-scale structure of the universe (LSS), void environments emerge as regions characterised by an average underdensity, hosting galaxies in their most pristine states, predominantly featuring bluer, disc-like galaxies, with active star formation \citep{2004ApJ...617...50R, 2005ApJ...624..571R, 2021ApJ...906...97F, 2023MNRAS.521..916R}. Conversely, higher density environments yield redder, elliptical galaxies, as predicted by the morphology-density relation \citep{1980ApJ...236..351D}. It is thus reasonable to conceive environment density as a potential driver for the suppression of star formation: as a matter of fact, mass and environment have been previously proposed as two main, independent mechanisms for galaxy quenching \citep{2010ApJ...721..193P}.
        
    The effect of extragalactic environment in local scales has been extensively studied, specially in terms of close pairs, compact groups, and merging systems. There is evidence of enhanced star formation in close pairs of galaxies, with the highest enhancement in pairs of galaxies of similar mass and a decrease towards higher separations in the sky plane \citep{2003MNRAS.346.1189L, 2008AJ....135.1877E, 2011MNRAS.412..591P, 2012MNRAS.426..549S}. However, \cite{2013MNRAS.433L..59P} show clear evidence of enhanced star formation in pairs up to ${\sim} 150 \, \text{kpc}$ separation, making use of both observational and simulation techniques. Regarding merger systems, \cite{2010MNRAS.407.1514E} show how the enhancement of post-merger star formation is higher in low-intermediate density environments, in addition to higher asymmetry, bluer bulges and higher bulge-to-total flux fractions. On the other hand, simulations focusing on central-satellite dynamics find that satellite galaxies suffer rapid quenching episodes after the infall into their host halos \citep{2013MNRAS.432..336W, 2023ApJ...954...98P}.
    
    Despite the extensive literature addressing the impact of local environment on galaxy evolution, the interplay between local and large-scale environments and their collective influence on galaxies remains an almost unexplored frontier. In \cite{2018MNRAS.481.2458D} it is discussed that galaxies in pairs are more likely to present enhanced star formation due to galaxy interactions \citep[in agreement with previous works, i.e.][]{1996ApJ...471..115B} than galaxies in triplets and groups, which are prone to suffer quenching due to the multiple interactions within the group. It is later discussed in \cite{2020MNRAS.493.1818D} that those effects may be caused, not only by the number of galaxies in the group, but by the implicit LSS. Other recent studies focus on different aspects related to this local and global interplay, such as properties (i.e. colour, star formation rate, and concentration) of void galaxies in groups \citep{2023MNRAS.521..916R}, brightest group galaxies \citep{2020A&A...639A..71K, 2024A&A...681A..91E}, compact groups \citep{2023MNRAS.520.6367T} or isolated triplets \citep{2023A&A...670A..63V}, in diverse LSS environments. All these works find that group galaxy properties are influenced by their position in the cosmic web, although how this influence acts on galaxies over time is still a missing piece. In this work, we make our contribution by looking further to the past through star formation histories (SFHs) in galaxies of different local and global environments.

    This paper is structured as follows: in Sect. \ref{sec:datasample}, we describe the data and samples used in this work, as well as the large and local scale environment classification of the galaxy sample. In Sect. \ref{sec:results}, we explore the SFHs of the different subsamples. We discuss the results in Sect. \ref{sec:discussion}. Finally, we present a summary of the work and our main conclusions in Sect. \ref{sec:conclusions}. Throughout the study, a cosmology with $\Omega_{\Lambda_0}=0.7$, $\Omega_{m_0}=0.3$, and $H_0=67.8\, \mathrm{km \, s^{-1} \, Mpc^{-1}}$ is considered.


\section{Data \& sample} \label{sec:datasample}

    The main sample contains 9384 galaxies within the redshift range $~{0.01<z<0.05}$ and stellar mass range ${\log{(M_\bigstar / \mathrm{M_\odot)}} > 9.0}$, with SFH data provided by \cite{2023Natur.619..269D}, as well as information about their LSS and local environment. In the following paragraphs we address the details regarding the determination of SFHs and the environmental characterisation.
    
    In \cite{2023Natur.619..269D}, the SFHs are estimated through the application of stellar spectrum models \citep{2016MNRAS.463.3409V} to spectra from Sloan Digital Sky Survey Data Release 16 \citep[SDSS DR16; ][]{2020ApJS..249....3A}. The extraction of stellar populations was facilitated using a combination of full-spectral fitting codes, namely the penalised pixel fitting \citep[PPXF; ][]{2017MNRAS.466..798C} code and the STEllar Content and Kinematics from high-resolution galactic spectra algorithm \citep[STECKMAP;][]{2006MNRAS.365...74O}. Our study starts off from a set of 10807 galaxies with computed SFHs \citep[987 in voids, 6463 in filaments and 3357 in clusters, according to ][]{2023Natur.619..269D}.

    In order to constrain the local environment, we make use of the galaxy catalogue given by \cite{2017A&A...602A.100T}. This catalogue contains 584449 galaxies from SDSS DR12 \citep{2011AJ....142...72E}, and provides information about galaxy groups and clusters; however, it is not restricted to galaxies in groups, and they provide a clean, large galaxy sample, suitable for this investigation into the local environment. Companion galaxies are searched for every galaxy within velocity differences ${\Delta v \leq 160 \, \text{km s}^{-1}}$ and sky-projected distances ${\Delta r_p \leq 0.45 \, \text{Mpc}}$, following the criteria for physical bounding presented in \cite{2015A&A...578A.110A}. They find evidence of physical bounding in isolated pairs and triplets within those conditions, hence these criteria outline a conservative approach to consider galaxies in groups. This search is carried out within the \cite{2017A&A...602A.100T} catalogue itself. In this study, galaxies are classified as either singlets or group members, regarding the absence or presence of a companion given the aforementioned conditions. Consequently, group galaxies are those with a number of neighbours $N_{\mathrm{neigh}} \geq 1$.

    The final dataset is obtained through the crossmatch of the 10807 galaxies with SFH data and the \cite{2017A&A...602A.100T} catalogue, for which we have determined local environment. We are left with 9862 galaxies, which we characterise with respect to their placement within the LSS of the Universe, following the methodology in \cite{2023Natur.619..269D}. We identify void galaxies by cross-referencing the dataset with a void galaxy catalogue \citep{2012MNRAS.421..926P}. We consider cluster galaxies those which are part of groups of 30 or more galaxies \citep{1989ApJS...70....1A} in the group catalogue by \cite{2017A&A...602A.100T}. All other galaxies are categorised as filament or wall galaxies, which we henceforth refer to as filament galaxies. We discard 242 galaxies which appear as void and cluster galaxies simultaneously according to the described criteria. 
    
    Lastly, as we aim to create a volume-limited sample to ensure a proper environment characterisation, we apply a stellar mass cut of ${\log{(M_\star / \mathrm{M_\odot})} > 9}$ on both our primary sample and the catalogue provided by \cite{2017A&A...602A.100T}, used for the local environment characterisation. We are left with the final dataset of 9384 galaxies, which comprises 913 galaxies in voids, 5388 in filaments, and 3083 in clusters. In terms of the local environment, 5303 galaxies are part a group ($N_\mathrm{neigh} \geq 1$), while 4081 galaxies are singlets. In Table \ref{tab:sample}, a breakdown of the galaxy sample is portrayed.

    \begin{table}[h]
    \caption{Summary of the environment classification of the galaxy sample.}
        \centering
          \begin{tabular}{llc}
        \hline\hline
        LSS & Local environment & Subsample size \\ \hline
        cluster & singlet &    534 \\
        & group member &   2549 \\ \hline
filament & singlet &   2897 \\
& group member &   2491 \\ \hline
void & singlet &    650 \\
& group member &    263 \\
        \hline
    \end{tabular}
    \label{tab:sample}
    \end{table}


\section{Results} \label{sec:results}

In this section, we describe the mass assembly look-back time distributions across different environments and explore how local and large-scale structure environments influence these patterns.

\subsection{Mass assembly look-back time distributions}

\begin{figure*}
\centering
\includegraphics[width=.47\textwidth]{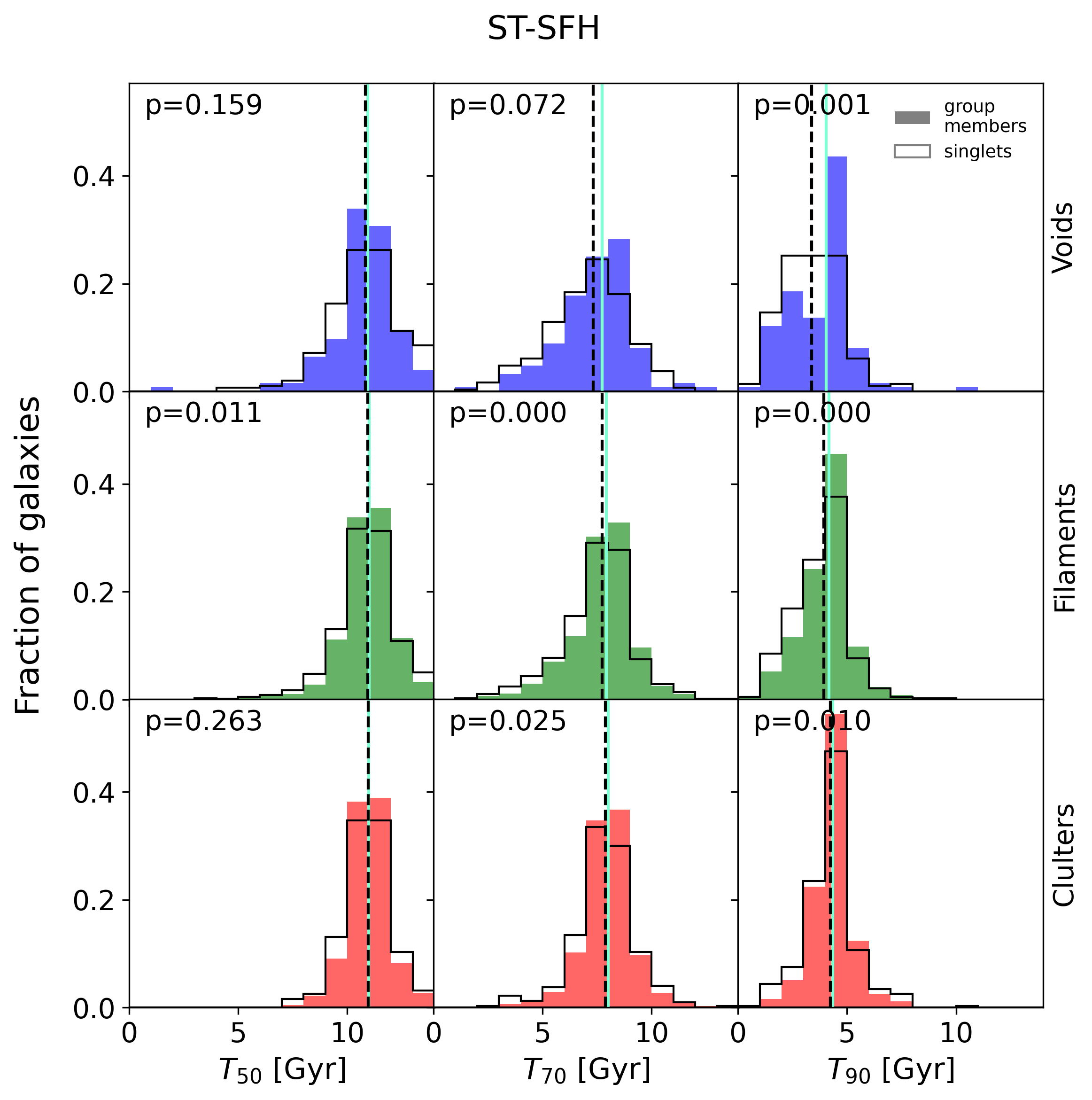}
\includegraphics[width=.47\textwidth]{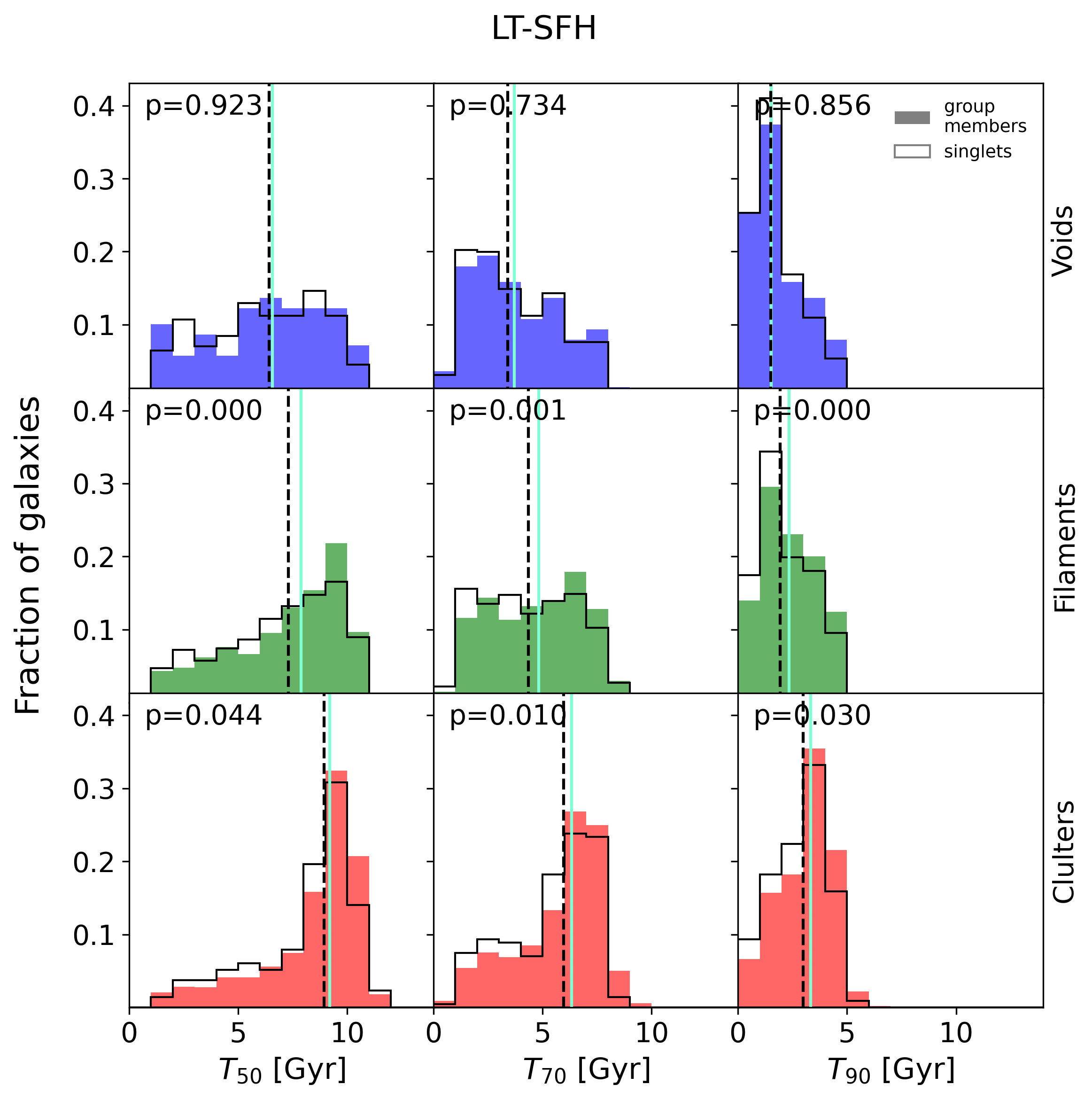}
 \caption{Stellar mass assembly look-back time distributions ($T_{50}$, $T_{70}$, and $T_{90}$) for singlets (black solid line) and galaxies in groups (colour-filled). \textit{On the left,} ST-SFH galaxies; \textit{on the right}, LT-SFH galaxies. The LSS environment is indicated on the right axis, and is colour-coded (voids in blue, filaments and walls in green, and clusters in red). The median of the distribution appears in solid cyan (group members) and dashed black (singlets) vertical lines. The values on the upper left corner correspond to the p-value from the Kolmogorov-Smirnov tests conducted between singlets and grouped galaxies distributions in each subpanel. All distributions contain over 150 galaxies.}
 \label{fig:lookbacktimes}
\end{figure*}

Mass assembly look-back time ($T_\%$) is defined as the span of time elapsed since a galaxy acquired a specific portion of its current stellar mass ($T_{50}$ for 50\%, $T_{70}$ for 70\%, and so on). Larger look-back times denote earlier times in the history of the Universe, and as a result, $T_{50}>T_{70}>T_{90}$. It is a representative parameter of the SFH and will be used consequently along this work. Moreover, galaxies are classified into two types regarding their SFH: galaxies with short-timescale SFH (ST-SFH) assemble most of their mass quickly and early, whereas long-timescale SFH (LT-SFH) galaxies assemble their mass steadily along cosmic time. According to \cite{2023Natur.619..269D}, this classification is independent of the LSS location.

Fig. \ref{fig:lookbacktimes} displays the look-back time distributions for both singlet and group galaxies, encompassing all LSS environments. In the left-hand panel, corresponding to ST-SFHs, the dissimilarity between the distributions of singlet and group galaxy samples becomes evident, as the mode of the distributions of group members reaches larger look-back times values, especially in regions of lower density and more recent time periods. We find the highest delay in median values in $T_{90}$ void galaxy distributions (0.8 Gyr), which decreases in $T_{70}$ and $T_{50}$ (0.6 Gyr and $<$0.2 Gyr, respectively). This distinction is no longer discernible in LT-SFHs (right-hand panel), where the distributions are broader and statistically similar for singlets and galaxies in groups. This outcome is substantiated by the results of the two-sample Kolmogorov–Smirnov tests conducted for each panel. In ST-SFH galaxies, only 2 out of 9 p-values are greater than 0.05 ($T_{50}$ distributions in filaments and clusters); however, in LT-SFH galaxies, only filament galaxies show p-values under 0.05. As a result, the null hypothesis (that the two samples are drawn from the same distribution) can be rejected in ST-SFH galaxies, and we can assume that singlets and galaxies in groups have inherently different assembly look-back time distributions, whereas this is not true for LT-SFH galaxies.

\subsection{Assembly times and stellar mass}

\begin{figure}
\centering
\includegraphics[width=0.485\textwidth]{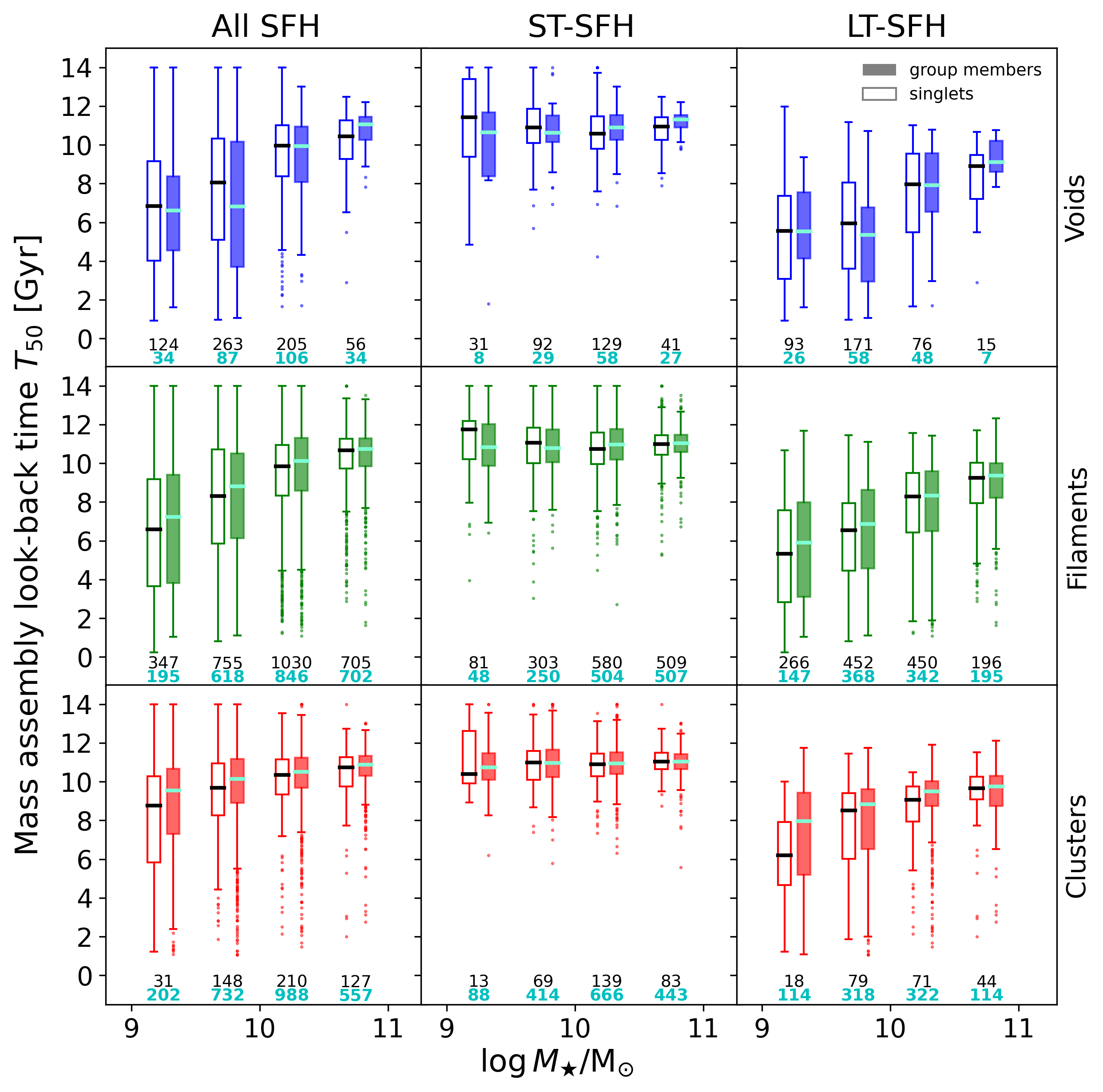}
\\ \vspace{.35cm}
\includegraphics[width=0.485\textwidth]{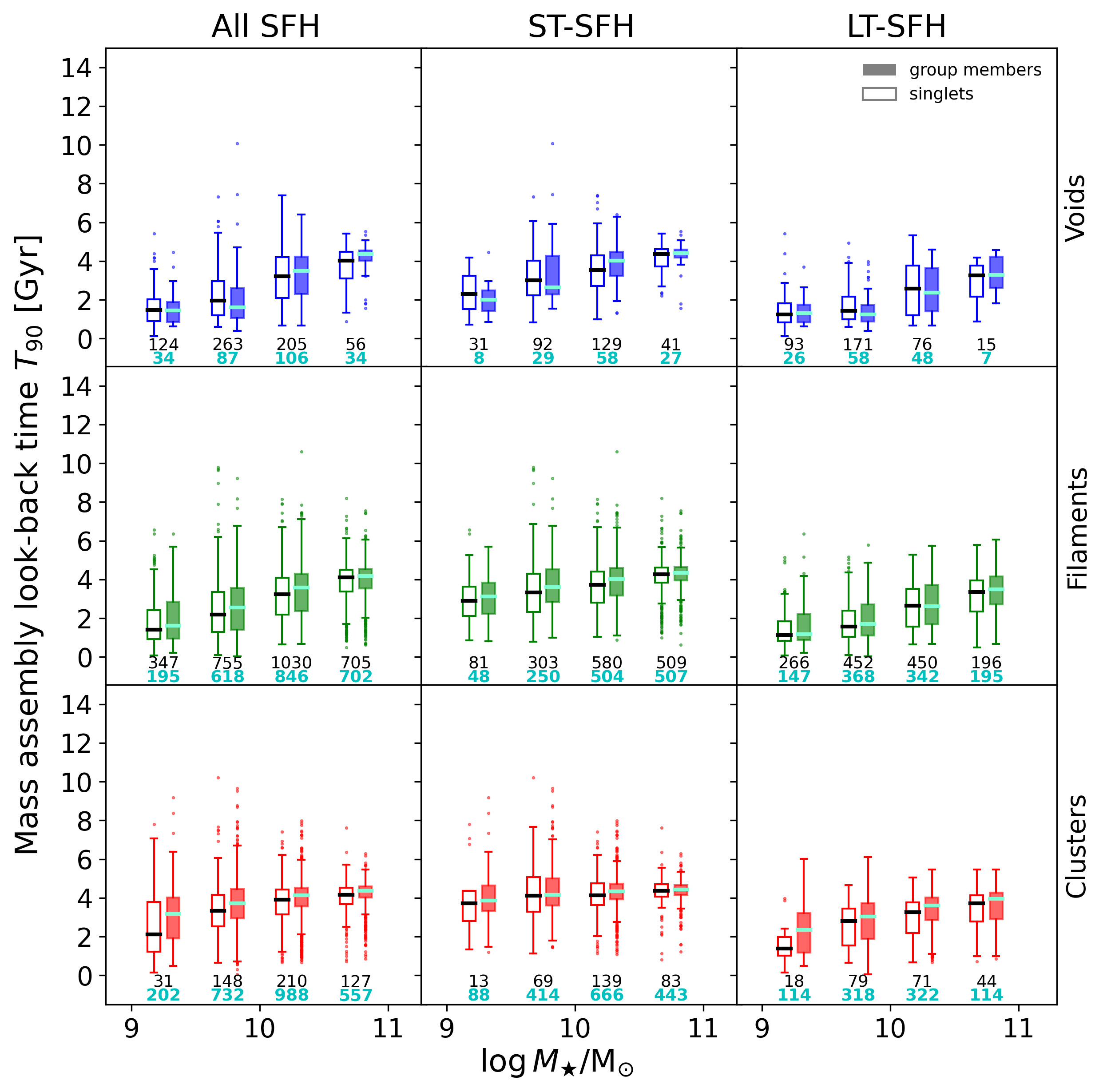}
\caption{Mass assembly look-back time distributions vs. stellar mass for singlets (empty boxes) and grouped galaxies (filled boxes) in voids (blue), filaments and walls (green) and clusters (red). \textit{In the left and right upper panels}, $T_{50}$ and $T_{70}$ distributions, respectively. \textit{In the lower panel}, $T_{90}$ distributions. The stellar mass bins are distributed from $10^{9}$ to $10^{11}\mathrm{M_\odot}$, in 0.5 dex intervals. The numbers in the lower end of each panel correspond to the size of the subsamples (black for singlets and cyan for galaxies in groups).}
\label{fig:lookbacktimes-mass}
\end{figure}

\begin{figure}
\centering
\includegraphics[width=0.48\textwidth]{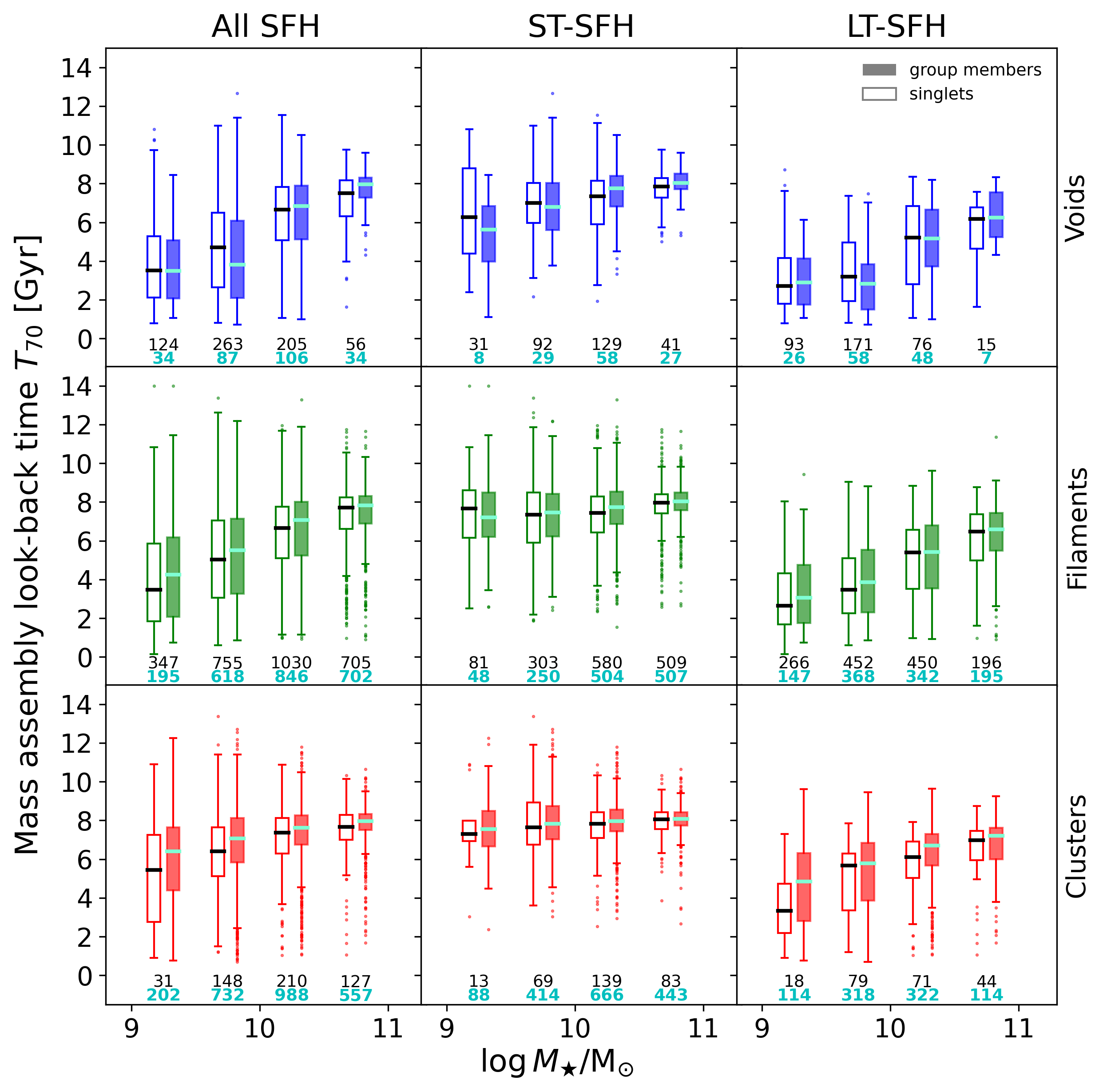}
\end{figure}

Fig. \ref{fig:lookbacktimes-mass} illustrates the look-back time distributions within stellar mass bins for various local and LSS environment subsamples. If all SFH types are considered, the distributions follow the trend of LT-SFHs, and appear generally widened to cover ST-SFHs distributions. Consistent with previous findings \citep{2023Natur.619..269D}, ST-SFH distributions are nearly constant with stellar mass, as opposed to LT-SFHs, in which lower look-back times are evident in the lower mass distributions. In ST-SFH galaxies, this result is found in cluster galaxy samples over all distributions ($T_{50}$, $T_{70}$, and $T_{90}$). However, ST-SFH void galaxies smoothly transition to resemble LT-SFH galaxies with increasing mass fraction: low mass galaxies display lower $T_{70}$ and $T_{90}$ than higher mass galaxies, especially in the case of galaxies in groups, in spite of the constant trend with mass seen in $T_{50}$ distributions. A similar behaviour is only mildly noticeable in filament galaxies.

The mass assembly look-back times distributions of singlets and groups exhibit both very similar behaviours, up to the point that they appear visually indistinguishable in ST-SFH galaxies. Notable discrepancies appear in LT-SFH galaxies within clusters, where it seems that group members may have assembled their stellar mass earlier than singlets, particularly in lower-mass distributions. These differences are not observed in void and filament galaxies with LT-SFH, although it is unclear whether the distributions are identical for singlet and group galaxies in those environments. Nevertheless, Fig. \ref{fig:lookbacktimes-mass} demonstrates that singlets and group galaxies do not exhibit a correlation between their SFHs and ST and LT-SFH types. Consequently, we can conclude that the local environment is not a primary driver of the observed SFH bimodality.

\begin{figure*}
\centering
    \includegraphics[width=.8\textwidth]{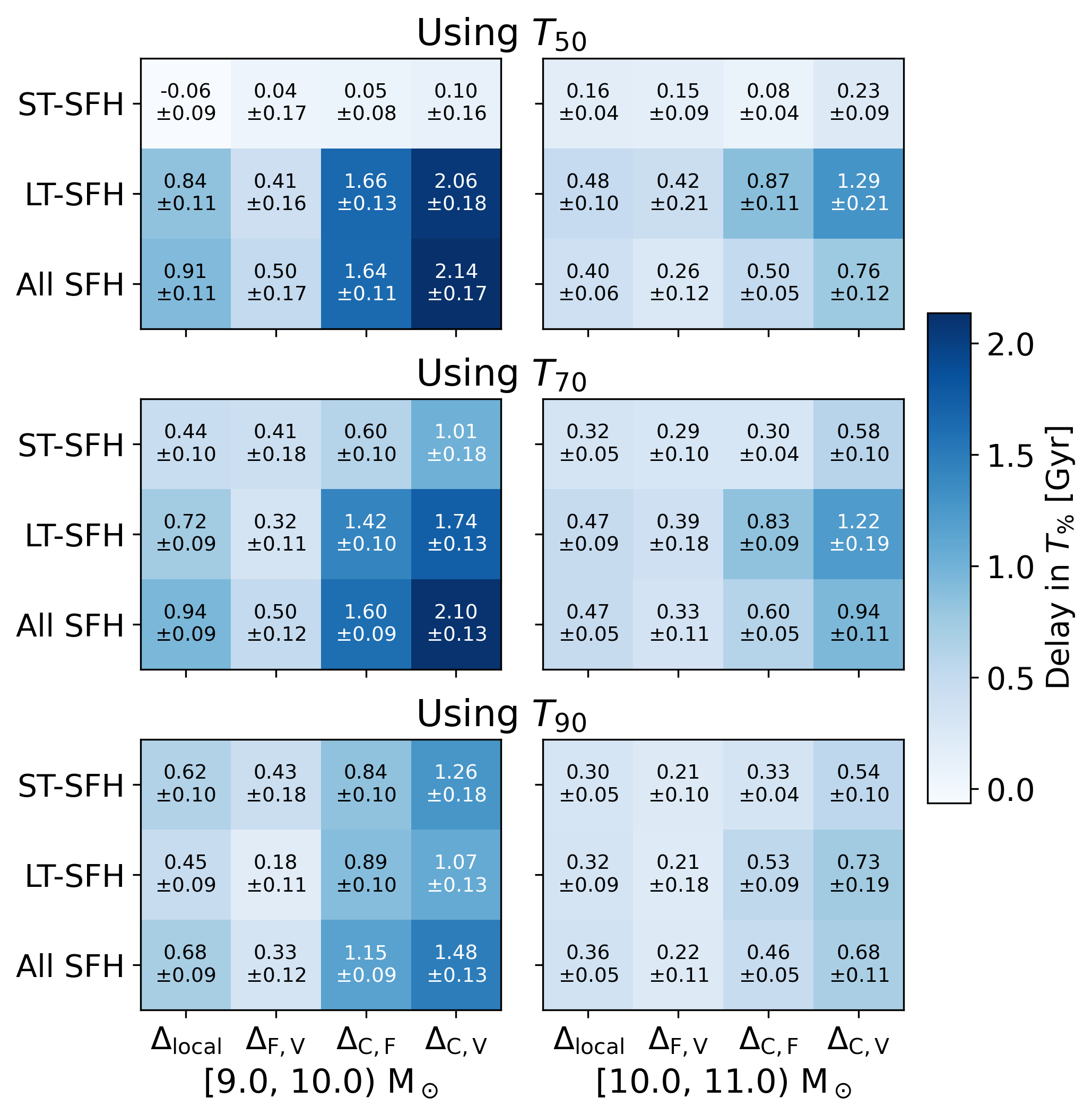}
    \caption{Delay in mean mass assembly look-back times caused by environment ($\Delta_{\mathrm{env}}$). The results using $T_{50}$, $T_{70}$ and $T_{90}$ appear in the first, second and third panels, reading from top to bottom. We consider galaxies of low ($10^{9}$- $10^{10}\mathrm{M_\odot}$) and high masses ($10^{10}$-$10^{11}\mathrm{M_\odot}$) on the left and right panels. Every column corresponds to a $\Delta_{\mathrm{env}}$ value, where C, F and V depict clusters, filaments and voids respectively, aiming to compare every LSS environment. $\Delta_{\mathrm{local}}$ corresponds to the difference between groups and singlets. Rows in every panel depict SFH types (short-timescale SFH, long-timescale SFH, or both altogether).}
    \label{fig:delta}
\end{figure*}

A parametrisation of the effect of local ($\Delta_{\mathrm{local}}$) and LSS environments ($\Delta_{\mathrm{C,V}}$, $\Delta_{\mathrm{F,V}}$, and $\Delta_{\mathrm{C,F}}$) on SFHs can be found in Fig. \ref{fig:delta}. For every two environments compared, these $\Delta_{\mathrm{env}}$ parameters are obtained by subtracting the mean $T_{\%}$ of the less dense to the denser environment. In the case of the LSS environments, all combinations are considered. The result is a quantification of the mean delay caused by environment density. According to these results, low mass galaxies are significantly more affected than high mass galaxies, although the delay caused by clusters with respect to any other LSS environment is also found in high mass galaxies. In general, these delays are the most significant (up to $\sim$2 Gyr), followed by the differences of groups with respect to singlets ($\Delta_{\mathrm{local}}<0.9$ Gyr). We find the lowest values when comparing voids and filaments ($\Delta_{\mathrm{F,V}}<0.6$ Gyr). Regarding SFH types, ST-SFH galaxies are affected in a similar manner by all environments when assembling 50\% of their mass, in contrast to later in time. LT-SFH galaxies show the largest delays, and impact greatly on the full sample, as previously seen in Fig. \ref{fig:lookbacktimes-mass}.

\section{Discussion} \label{sec:discussion}

\subsection{Mass assembly look-back time distributions}

The mass assembly history in different local and global environments can be inferred from the look-back time distributions in Fig. \ref{fig:lookbacktimes}. In galaxies with ST-SFH, the median of the $T_{50}$ distributions takes the value $\sim$11 Gyr, independently of the kind of local or LSS environment. At those early times ($\sim$3 Gyr after the Big Bang), the density contrasts between the forthcoming LSS were minor. Consequently, not finding significant differences both in local and LSS environments could be expected, as it is a reflection of the more homogeneous environment galaxies were embedded in during that early cosmological time. Differences appear in the $T_{70}$ distributions, in which voids show wider distributions, and singlets form the 70\% of their mass later than galaxies in groups, and therefore evolve slower. This effect is accentuated in the $T_{90}$ distributions, in which there is a $\sim$2 Gyr difference between the mode of the singlets and group members distributions in voids. These results point towards an accelerated evolutionary scenario for ST-SFH galaxies in groups. However, the physical scenario depends heavily on the mass of the galaxy and the characteristics of the group (e.g. number of members, compactness, etc) as supported by previous studies \citep{2018MNRAS.481.2458D, 2020MNRAS.493.1818D}. 

On the right panel of Fig. \ref{fig:lookbacktimes} we find galaxies with LT-SFH type. These galaxies assemble their mass progressively, exhibiting wider distributions in look-back time. Void galaxies assemble their mass slower than cluster galaxies, and therefore the maxima of the distributions are shifted towards more recent cosmic times in less dense environments. In LT-SFH galaxies, the effect of local environment does not emerge as clearly as in ST-SFH galaxies. 

\subsection{Assembly times and stellar mass}

\begin{figure}
    \centering
    \includegraphics[width=.48\textwidth]{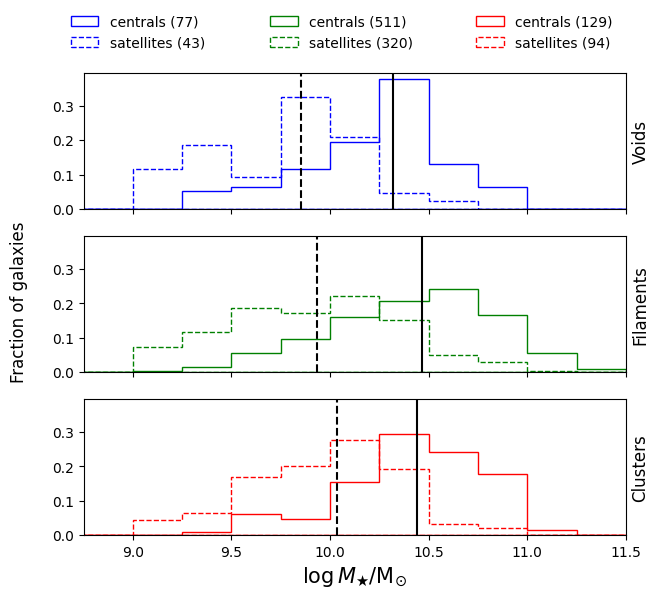}
    \caption{Stellar mass distributions of galaxies in galaxy pairs. \textit{From top to bottom}, blue, green and red histograms depict samples in voids, filaments and clusters. Dashed and solid lines correspond to the median values of the distributions of satellites and centrals, respectively. The numbers in the legend correspond to the subsample sizes.}
    \label{fig:mass_cn}
\end{figure}

To see how the masses in central and satellite galaxies are distributed in the different environments, we create a subsample of galaxies in pairs in our catalogue for companion search, and identify the central galaxy in each pair as the one with the largest stellar mass. In our main sample of 9384 galaxies, 1284 galaxies (13.4\%) are part of pairs. Fig. \ref{fig:mass_cn} illustrates the mass distributions of central and satellite galaxies, corresponding to the 1174 galaxies in pairs with both the central and satellite galaxy in the same LSS environment. It can be seen how dense environments contain more massive galaxies, with less difference in the distributions of central and satellite galaxies, in opposition to void galaxies. Hence, as illustrated in Fig. \ref{fig:lookbacktimes-mass}, satellites undergo more effective quenching in high-density environments. This is noticeable by their large look-back times compared to singlets in low-mass galaxies. These findings align with prior research, both in simulations and observations, highlighting a correlation between increased quenching efficiency and the mass of the host halo \citep{2019MNRAS.484.1702P, 2021MNRAS.502.4457G, 2023ApJ...954...98P}.

Recent works in simulations \citep{2022MNRAS.517..712R, 2024MNRAS.tmp..184R} find that void galaxies tend to experience later mergers than galaxies in denser environments, leading to mass accretion in later cosmic times. These results align with later look-back times in voids, and motivates further investigation into the SFHs of mergers in voids. Additionally, they find that central galaxies do not exhibit many differences in between LSS environments, in agreement with our results regarding high mass galaxies, along with other studies on the brightest group galaxies \citep{2020A&A...639A..71K, 2024A&A...681A..91E}. 

Ultimately, it can be understood that the SFH of galaxies is influenced by their local environment, although the influence of the LSS of the Universe can be larger in the densest structures. In any case, the nature of these influences is conditioned by stellar mass. The root of the differences in between global environments is yet unclear, although several mechanisms have been proposed in previous studies, such as the prevalence of the cold accretion mode of gas \citep{2005MNRAS.363....2K}. \cite{2023Natur.619..269D} offers additional discussion on this topic. It is noteworthy that the bimodal nature of SFHs persists regardless of the environmental context as addressed in this work.


\section{Summary and conclusions} \label{sec:conclusions}

We characterise the local and large scale environment of 9384 galaxies, with the aim of analysing SFHs and possible evolutionary paths driven by the local environment. The sample is divided in three large scale structures (voids, walls \& filaments, and clusters) and two possible local environment classifications (singlets or group members) regarding the presence or absence of companions within the reach of physical bounding.

The following constitute our main results:
\begin{enumerate}    
    \item We address the bimodality in SFHs found by \cite{2023Natur.619..269D}. As defined in the aforementioned work, and explained in Sect. \ref{sec:results}, there appear to exist two main types of SFHs: ST-SFHs and LT-SFHs. This bimodality emerges clearly in both the group and singlet samples. Therefore, the local environment is not the main driving mechanism of the SFH bimodality.

    \item Mass assembly look-back time distributions indicate that ST-SFH galaxies in groups assemble their stellar mass earlier than singlets. When considering look-back times with respect to stellar mass bins, we also find this result for LT-SFH galaxies. These results point towards a common evolutionary background for galaxies in groups. This scenario is likely to alter SFHs in different ways in satellite and central galaxies, suppressing or inducing star formation.

    \item Environment influences the SFHs of galaxies. The LSS environment can cause a delay in stellar mass assembly of up to $\sim$2 Gyr, $<$1 Gyr in the case of local environment. The effect of both environments is more significant in less massive galaxies, and in LT-SFH galaxies. There are no large differences between the SFHs of void and filament galaxies, finding the largest differences when comparing galaxies inhabiting clusters to others populating voids or filaments.
\end{enumerate}

Given the dynamical nature of the large-scale structure of the universe, works like this one, which rely on observational data to trail galaxies to their birth, can gain invaluable benefits from the support of hydrodynamical simulations. In these simulations, tracing the environment of galaxies over cosmic time may become possible, a task that is inherently unfeasible for observations. Moreover, we require a more comprehensive understanding of the physical mechanisms that drive star formation in galaxies to fully interpret the presented results.

In spite of these limitations, this study has, for the first time, demonstrated that statistical discrepancies emerge in the star formation histories of galaxies that inhabit different local environments, as was similarly found for large-scale environments in \cite{2023Natur.619..269D}.


\begin{acknowledgements}
We are grateful to the anonymous referee, whose suggestions truly helped the authors to improve this manuscript. We acknowledge financial support from the research project PRE2021-098736 funded by MCIN/AEI/10.13039/501100011033 and FSE+. We acknowledge financial support by the research projects AYA2017-84897-P, PID2020-113689GB-I00, and PID2020-114414GB-I00, financed by MCIN/AEI/10.13039/501100011033, the project A-FQM-510-UGR20 financed from FEDER/Junta de Andalucía-Consejería de Transformación Económica, Industria, Conocimiento y Universidades/Proyecto and by the grants P20\_00334 and FQM108, financed by the Junta de Andalucía (Spain). M.A-F. acknowledges support from the Emergia program (EMERGIA20\_38888) from Consejer\'ia de Universidad, Investigaci\'on e Innovaci\'on de la Junta de Andaluc\'ia. S.D.P. acknowledges financial support from Juan de la Cierva Formaci\'on fellowship (FJC2021-047523-I) financed by MCIN/AEI/10.13039/501100011033 and by the European Union `NextGenerationEU'/PRTR, Ministerio de Econom\'ia y Competitividad under grants PID2019-107408GB-C44, PID2022-136598NB-C32, and is grateful to the Natural Sciences and Engineering Research Council of Canada, the Fonds de Recherche du Qu\'ebec, and the Canada Foundation for Innovation for funding. T.R.L. acknowledges support from Juan de la Cierva fellowship (IJC2020-043742-I). R.G.B., R.G.D., and A.C., acknowledge financial support from the Severo Ochoa grant CEX2021-001131-S funded by MCIN/AEI/ 10.13039/501100011033 and PID2022-141755NB-I00 and PID2019-109067GB-I00. J.F-B. acknowledges support from the PID2022-140869NB-I00 grant from the Spanish Ministry of Science and Innovation.

This research made use of Astropy, a community-developed core Python (\url{http://www.python.org}) package for astronomy \citep{2022ApJ...935..167A}; ipython \citep{PER-GRA:2007}; matplotlib \citep{Hunter:2007}; SciPy, a collection of open-source software for scientific computing in Python \citep{2020SciPy-NMeth}; and NumPy, a structure for efficient numerical computation \citep{harris2020array}. Funding for the SDSS and SDSS-II has been provided by the Alfred P. Sloan Foundation, the Participating Institutions, the National Science Foundation, the U.S. Department of Energy, the National Aeronautics and Space Administration, the Japanese Monbukagakusho, the Max Planck Society, and the Higher Education Funding Council for England. The SDSS Web Site is \url{http://www.sdss.org/}.
\end{acknowledgements}

\bibliographystyle{aa} 
\bibliography{paper} 

\begin{thebibliography}{41}
\expandafter\ifx\csname natexlab\endcsname\relax\def\natexlab#1{#1}\fi

\bibitem[{{Abell} {et~al.}(1989){Abell}, {Corwin}, \&
  {Olowin}}]{1989ApJS...70....1A}
{Abell}, G.~O., {Corwin}, Harold~G., J., \& {Olowin}, R.~P. 1989, \apjs, 70, 1

\bibitem[{{Ahumada} {et~al.}(2020){Ahumada}, {Allende Prieto}, {Almeida},
  {Anders}, {Anderson}, {Andrews}, {Anguiano}, {Arcodia}, {Armengaud},
  {Aubert}, {Avila}, {Avila-Reese}, {Badenes}, {Balland}, {Barger},
  {Barrera-Ballesteros}, {Basu}, {Bautista}, {Beaton}, {Beers}, {Benavides},
  {Bender}, {Bernardi}, {Bershady}, {Beutler}, {Bidin}, {Bird}, {Bizyaev},
  {Blanc}, {Blanton}, {Boquien}, {Borissova}, {Bovy}, {Brandt}, {Brinkmann},
  {Brownstein}, {Bundy}, {Bureau}, {Burgasser}, {Burtin}, {Cano-D{\'\i}az},
  {Capasso}, {Cappellari}, {Carrera}, {Chabanier}, {Chaplin}, {Chapman},
  {Cherinka}, {Chiappini}, {Doohyun Choi}, {Chojnowski}, {Chung}, {Clerc},
  {Coffey}, {Comerford}, {Comparat}, {da Costa}, {Cousinou}, {Covey}, {Crane},
  {Cunha}, {Ilha}, {Dai}, {Damsted}, {Darling}, {Davidson}, {Davies}, {Dawson},
  {De}, {de la Macorra}, {De Lee}, {Queiroz}, {Deconto Machado}, {de la Torre},
  {Dell'Agli}, {du Mas des Bourboux}, {Diamond-Stanic}, {Dillon}, {Donor},
  {Drory}, {Duckworth}, {Dwelly}, {Ebelke}, {Eftekharzadeh}, {Davis Eigenbrot},
  {Elsworth}, {Eracleous}, {Erfanianfar}, {Escoffier}, {Fan}, {Farr},
  {Fern{\'a}ndez-Trincado}, {Feuillet}, {Finoguenov}, {Fofie},
  {Fraser-McKelvie}, {Frinchaboy}, {Fromenteau}, {Fu}, {Galbany}, {Garcia},
  {Garc{\'\i}a-Hern{\'a}ndez}, {Garma Oehmichen}, {Ge}, {Geimba Maia},
  {Geisler}, {Gelfand}, {Goddy}, {Gonzalez-Perez}, {Grabowski}, {Green},
  {Grier}, {Guo}, {Guy}, {Harding}, {Hasselquist}, {Hawken}, {Hayes}, {Hearty},
  {Hekker}, {Hogg}, {Holtzman}, {Horta}, {Hou}, {Hsieh}, {Huber}, {Hunt}, {Ider
  Chitham}, {Imig}, {Jaber}, {Jimenez Angel}, {Johnson}, {Jones},
  {J{\"o}nsson}, {Jullo}, {Kim}, {Kinemuchi}, {Kirkpatrick}, {Kite}, {Klaene},
  {Kneib}, {Kollmeier}, {Kong}, {Kounkel}, {Krishnarao}, {Lacerna}, {Lan},
  {Lane}, {Law}, {Le Goff}, {Leung}, {Lewis}, {Li}, {Lian}, {Lin}, {Long},
  {Longa-Pe{\~n}a}, {Lundgren}, {Lyke}, {Mackereth}, {MacLeod}, {Majewski},
  {Manchado}, {Maraston}, {Martini}, {Masseron}, {Masters}, {Mathur},
  {McDermid}, {Merloni}, {Merrifield}, {M{\'e}sz{\'a}ros}, {Miglio}, {Minniti},
  {Minsley}, {Miyaji}, {Mohammad}, {Mosser}, {Mueller}, {Muna},
  {Mu{\~n}oz-Guti{\'e}rrez}, {Myers}, {Nadathur}, {Nair}, {Nandra}, {Correa do
  Nascimento}, {Nevin}, {Newman}, {Nidever}, {Nitschelm}, {Noterdaeme},
  {O'Connell}, {Olmstead}, {Oravetz}, {Oravetz}, {Osorio}, {Pace}, {Padilla},
  {Palanque-Delabrouille}, {Palicio}, {Pan}, {Pan}, {Parker}, {Paviot},
  {Peirani}, {Ram{\'r}ez}, {Penny}, {Percival}, {Perez-Fournon},
  {P{\'e}rez-R{\`a}fols}, {Petitjean}, {Pieri}, {Pinsonneault}, {Poovelil},
  {Povick}, {Prakash}, {Price-Whelan}, {Raddick}, {Raichoor}, {Ray}, {Rembold},
  {Rezaie}, {Riffel}, {Riffel}, {Rix}, {Robin}, {Roman-Lopes},
  {Rom{\'a}n-Z{\'u}{\~n}iga}, {Rose}, {Ross}, {Rossi}, {Rowlands}, {Rubin},
  {Salvato}, {S{\'a}nchez}, {S{\'a}nchez-Menguiano}, {S{\'a}nchez-Gallego},
  {Sayres}, {Schaefer}, {Schiavon}, {Schimoia}, {Schlafly}, {Schlegel},
  {Schneider}, {Schultheis}, {Schwope}, {Seo}, {Serenelli}, {Shafieloo},
  {Shamsi}, {Shao}, {Shen}, {Shetrone}, {Shirley}, {Silva Aguirre}, {Simon},
  {Skrutskie}, {Slosar}, {Smethurst}, {Sobeck}, {Sodi}, {Souto}, {Stark},
  {Stassun}, {Steinmetz}, {Stello}, {Stermer}, {Storchi-Bergmann},
  {Streblyanska}, {Stringfellow}, {Stutz}, {Su{\'a}rez}, {Sun},
  {Taghizadeh-Popp}, {Talbot}, {Tayar}, {Thakar}, {Theriault}, {Thomas},
  {Thomas}, {Tinker}, {Tojeiro}, {Toledo}, {Tremonti}, {Troup}, {Tuttle},
  {Unda-Sanzana}, {Valentini}, {Vargas-Gonz{\'a}lez}, {Vargas-Maga{\~n}a},
  {V{\'a}zquez-Mata}, {Vivek}, {Wake}, {Wang}, {Weaver}, {Weijmans}, {Wild},
  {Wilson}, {Wilson}, {Wolthuis}, {Wood-Vasey}, {Yan}, {Yang}, {Y{\`e}che},
  {Zamora}, {Zarrouk}, {Zasowski}, {Zhang}, {Zhao}, {Zhao}, {Zheng}, {Zheng},
  {Zhu}, \& {Zou}}]{2020ApJS..249....3A}
{Ahumada}, R., {Allende Prieto}, C., {Almeida}, A., {et~al.} 2020, \apjs, 249,
  3

\bibitem[{{Argudo-Fern{\'a}ndez} {et~al.}(2015){Argudo-Fern{\'a}ndez},
  {Verley}, {Bergond}, {Duarte Puertas}, {Ramos Carmona}, {Sabater},
  {Fern{\'a}ndez Lorenzo}, {Espada}, {Sulentic}, {Ruiz}, \&
  {Leon}}]{2015A&A...578A.110A}
{Argudo-Fern{\'a}ndez}, M., {Verley}, S., {Bergond}, G., {et~al.} 2015, \aap,
  578, A110

\bibitem[{{Astropy Collaboration} {et~al.}(2022){Astropy Collaboration},
  {Price-Whelan}, {Lim}, {Earl}, {Starkman}, {Bradley}, {Shupe}, {Patil},
  {Corrales}, {Brasseur}, {N{\"o}the}, {Donath}, {Tollerud}, {Morris},
  {Ginsburg}, {Vaher}, {Weaver}, {Tocknell}, {Jamieson}, {van Kerkwijk},
  {Robitaille}, {Merry}, {Bachetti}, {G{\"u}nther}, {Aldcroft},
  {Alvarado-Montes}, {Archibald}, {B{\'o}di}, {Bapat}, {Barentsen},
  {Baz{\'a}n}, {Biswas}, {Boquien}, {Burke}, {Cara}, {Cara}, {Conroy},
  {Conseil}, {Craig}, {Cross}, {Cruz}, {D'Eugenio}, {Dencheva}, {Devillepoix},
  {Dietrich}, {Eigenbrot}, {Erben}, {Ferreira}, {Foreman-Mackey}, {Fox},
  {Freij}, {Garg}, {Geda}, {Glattly}, {Gondhalekar}, {Gordon}, {Grant},
  {Greenfield}, {Groener}, {Guest}, {Gurovich}, {Handberg}, {Hart},
  {Hatfield-Dodds}, {Homeier}, {Hosseinzadeh}, {Jenness}, {Jones}, {Joseph},
  {Kalmbach}, {Karamehmetoglu}, {Ka{\l}uszy{\'n}ski}, {Kelley}, {Kern},
  {Kerzendorf}, {Koch}, {Kulumani}, {Lee}, {Ly}, {Ma}, {MacBride}, {Maljaars},
  {Muna}, {Murphy}, {Norman}, {O'Steen}, {Oman}, {Pacifici}, {Pascual},
  {Pascual-Granado}, {Patil}, {Perren}, {Pickering}, {Rastogi}, {Roulston},
  {Ryan}, {Rykoff}, {Sabater}, {Sakurikar}, {Salgado}, {Sanghi}, {Saunders},
  {Savchenko}, {Schwardt}, {Seifert-Eckert}, {Shih}, {Jain}, {Shukla}, {Sick},
  {Simpson}, {Singanamalla}, {Singer}, {Singhal}, {Sinha}, {Sip{\H{o}}cz},
  {Spitler}, {Stansby}, {Streicher}, {{\v{S}}umak}, {Swinbank}, {Taranu},
  {Tewary}, {Tremblay}, {de Val-Borro}, {Van Kooten}, {Vasovi{\'c}}, {Verma},
  {de Miranda Cardoso}, {Williams}, {Wilson}, {Winkel}, {Wood-Vasey}, {Xue},
  {Yoachim}, {Zhang}, {Zonca}, \& {Astropy Project
  Contributors}}]{2022ApJ...935..167A}
{Astropy Collaboration}, {Price-Whelan}, A.~M., {Lim}, P.~L., {et~al.} 2022,
  \apj, 935, 167

\bibitem[{{Barnes} \& {Hernquist}(1996)}]{1996ApJ...471..115B}
{Barnes}, J.~E. \& {Hernquist}, L. 1996, \apj, 471, 115

\bibitem[{{Cappellari}(2017)}]{2017MNRAS.466..798C}
{Cappellari}, M. 2017, \mnras, 466, 798

\bibitem[{{Dom{\'\i}nguez-G{\'o}mez} {et~al.}(2023){Dom{\'\i}nguez-G{\'o}mez},
  {P{\'e}rez}, {Ruiz-Lara}, {Peletier}, {S{\'a}nchez-Bl{\'a}zquez},
  {Lisenfeld}, {Falc{\'o}n-Barroso}, {Alc{\'a}zar-Laynez},
  {Argudo-Fern{\'a}ndez}, {Bl{\'a}zquez-Calero}, {Courtois}, {Duarte Puertas},
  {Espada}, {Florido}, {Garc{\'\i}a-Benito}, {Jim{\'e}nez}, {Kreckel},
  {Rela{\~n}o}, {S{\'a}nchez-Menguiano}, {van der Hulst}, {van de Weygaert},
  {Verley}, \& {Zurita}}]{2023Natur.619..269D}
{Dom{\'\i}nguez-G{\'o}mez}, J., {P{\'e}rez}, I., {Ruiz-Lara}, T., {et~al.}
  2023, \nat, 619, 269

\bibitem[{{Dressler}(1980)}]{1980ApJ...236..351D}
{Dressler}, A. 1980, \apj, 236, 351

\bibitem[{{Duplancic} {et~al.}(2018){Duplancic}, {Coldwell}, {Alonso}, \&
  {Lambas}}]{2018MNRAS.481.2458D}
{Duplancic}, F., {Coldwell}, G.~V., {Alonso}, S., \& {Lambas}, D.~G. 2018,
  \mnras, 481, 2458

\bibitem[{{Duplancic} {et~al.}(2020){Duplancic}, {D{\'a}vila-Kurb{\'a}n},
  {Coldwell}, {Alonso}, \& {Galdeano}}]{2020MNRAS.493.1818D}
{Duplancic}, F., {D{\'a}vila-Kurb{\'a}n}, F., {Coldwell}, G.~V., {Alonso}, S.,
  \& {Galdeano}, D. 2020, \mnras, 493, 1818

\bibitem[{{Einasto} {et~al.}(2024){Einasto}, {Einasto}, {Tenjes}, {Korhonen},
  {Kipper}, {Tempel}, {Liivam{\"a}gi}, \&
  {Hein{\"a}m{\"a}ki}}]{2024A&A...681A..91E}
{Einasto}, M., {Einasto}, J., {Tenjes}, P., {et~al.} 2024, \aap, 681, A91

\bibitem[{{Eisenstein} {et~al.}(2011){Eisenstein}, {Weinberg}, {Agol},
  {Aihara}, {Allende Prieto}, {Anderson}, {Arns}, {Aubourg}, {Bailey},
  {Balbinot}, {Barkhouser}, {Beers}, {Berlind}, {Bickerton}, {Bizyaev},
  {Blanton}, {Bochanski}, {Bolton}, {Bosman}, {Bovy}, {Brandt}, {Breslauer},
  {Brewington}, {Brinkmann}, {Brown}, {Brownstein}, {Burger}, {Busca},
  {Campbell}, {Cargile}, {Carithers}, {Carlberg}, {Carr}, {Chang}, {Chen},
  {Chiappini}, {Comparat}, {Connolly}, {Cortes}, {Croft}, {Cunha}, {da Costa},
  {Davenport}, {Dawson}, {De Lee}, {Porto de Mello}, {de Simoni}, {Dean},
  {Dhital}, {Ealet}, {Ebelke}, {Edmondson}, {Eiting}, {Escoffier}, {Esposito},
  {Evans}, {Fan}, {Femen{\'\i}a Castell{\'a}}, {Dutra Ferreira}, {Fitzgerald},
  {Fleming}, {Font-Ribera}, {Ford}, {Frinchaboy}, {Garc{\'\i}a P{\'e}rez},
  {Gaudi}, {Ge}, {Ghezzi}, {Gillespie}, {Gilmore}, {Girardi}, {Gott}, {Gould},
  {Grebel}, {Gunn}, {Hamilton}, {Harding}, {Harris}, {Hawley}, {Hearty},
  {Hennawi}, {Gonz{\'a}lez Hern{\'a}ndez}, {Ho}, {Hogg}, {Holtzman},
  {Honscheid}, {Inada}, {Ivans}, {Jiang}, {Jiang}, {Johnson}, {Jordan},
  {Jordan}, {Kauffmann}, {Kazin}, {Kirkby}, {Klaene}, {Knapp}, {Kneib},
  {Kochanek}, {Koesterke}, {Kollmeier}, {Kron}, {Lampeitl}, {Lang}, {Lawler},
  {Le Goff}, {Lee}, {Lee}, {Leisenring}, {Lin}, {Liu}, {Long}, {Loomis},
  {Lucatello}, {Lundgren}, {Lupton}, {Ma}, {Ma}, {MacDonald}, {Mack},
  {Mahadevan}, {Maia}, {Majewski}, {Makler}, {Malanushenko}, {Malanushenko},
  {Mandelbaum}, {Maraston}, {Margala}, {Maseman}, {Masters}, {McBride},
  {McDonald}, {McGreer}, {McMahon}, {Mena Requejo}, {M{\'e}nard},
  {Miralda-Escud{\'e}}, {Morrison}, {Mullally}, {Muna}, {Murayama}, {Myers},
  {Naugle}, {Neto}, {Nguyen}, {Nichol}, {Nidever}, {O'Connell}, {Ogando},
  {Olmstead}, {Oravetz}, {Padmanabhan}, {Paegert}, {Palanque-Delabrouille},
  {Pan}, {Pandey}, {Parejko}, {P{\^a}ris}, {Pellegrini}, {Pepper}, {Percival},
  {Petitjean}, {Pfaffenberger}, {Pforr}, {Phleps}, {Pichon}, {Pieri}, {Prada},
  {Price-Whelan}, {Raddick}, {Ramos}, {Reid}, {Reyle}, {Rich}, {Richards},
  {Rieke}, {Rieke}, {Rix}, {Robin}, {Rocha-Pinto}, {Rockosi}, {Roe},
  {Rollinde}, {Ross}, {Ross}, {Rossetto}, {S{\'a}nchez}, {Santiago}, {Sayres},
  {Schiavon}, {Schlegel}, {Schlesinger}, {Schmidt}, {Schneider}, {Sellgren},
  {Shelden}, {Sheldon}, {Shetrone}, {Shu}, {Silverman}, {Simmerer}, {Simmons},
  {Sivarani}, {Skrutskie}, {Slosar}, {Smee}, {Smith}, {Snedden}, {Stassun},
  {Steele}, {Steinmetz}, {Stockett}, {Stollberg}, {Strauss}, {Szalay},
  {Tanaka}, {Thakar}, {Thomas}, {Tinker}, {Tofflemire}, {Tojeiro}, {Tremonti},
  {Vargas Maga{\~n}a}, {Verde}, {Vogt}, {Wake}, {Wan}, {Wang}, {Weaver},
  {White}, {White}, {Wilson}, {Wisniewski}, {Wood-Vasey}, {Yanny}, {Yasuda},
  {Y{\`e}che}, {York}, {Young}, {Zasowski}, {Zehavi}, \&
  {Zhao}}]{2011AJ....142...72E}
{Eisenstein}, D.~J., {Weinberg}, D.~H., {Agol}, E., {et~al.} 2011, \aj, 142, 72

\bibitem[{{Ellison} {et~al.}(2008){Ellison}, {Patton}, {Simard}, \&
  {McConnachie}}]{2008AJ....135.1877E}
{Ellison}, S.~L., {Patton}, D.~R., {Simard}, L., \& {McConnachie}, A.~W. 2008,
  \aj, 135, 1877

\bibitem[{{Ellison} {et~al.}(2010){Ellison}, {Patton}, {Simard}, {McConnachie},
  {Baldry}, \& {Mendel}}]{2010MNRAS.407.1514E}
{Ellison}, S.~L., {Patton}, D.~R., {Simard}, L., {et~al.} 2010, \mnras, 407,
  1514

\bibitem[{{Florez} {et~al.}(2021){Florez}, {Berlind}, {Kannappan}, {Stark},
  {Eckert}, {Calderon}, {Moffett}, {Campbell}, \&
  {Sinha}}]{2021ApJ...906...97F}
{Florez}, J., {Berlind}, A.~A., {Kannappan}, S.~J., {et~al.} 2021, \apj, 906,
  97

\bibitem[{{Gallazzi} {et~al.}(2021){Gallazzi}, {Pasquali}, {Zibetti}, \&
  {Barbera}}]{2021MNRAS.502.4457G}
{Gallazzi}, A.~R., {Pasquali}, A., {Zibetti}, S., \& {Barbera}, F.~L. 2021,
  \mnras, 502, 4457

\bibitem[{Harris {et~al.}(2020)Harris, Millman, van~der Walt, Gommers,
  Virtanen, Cournapeau, Wieser, Taylor, Berg, Smith, Kern, Picus, Hoyer, van
  Kerkwijk, Brett, Haldane, del R{\'{i}}o, Wiebe, Peterson,
  G{\'{e}}rard-Marchant, Sheppard, Reddy, Weckesser, Abbasi, Gohlke, \&
  Oliphant}]{harris2020array}
Harris, C.~R., Millman, K.~J., van~der Walt, S.~J., {et~al.} 2020, Nature, 585,
  357

\bibitem[{Hunter(2007)}]{Hunter:2007}
Hunter, J.~D. 2007, Computing in Science \& Engineering, 9, 90

\bibitem[{{Kere{\v{s}}} {et~al.}(2005){Kere{\v{s}}}, {Katz}, {Weinberg}, \&
  {Dav{\'e}}}]{2005MNRAS.363....2K}
{Kere{\v{s}}}, D., {Katz}, N., {Weinberg}, D.~H., \& {Dav{\'e}}, R. 2005,
  \mnras, 363, 2

\bibitem[{{Kuutma} {et~al.}(2020){Kuutma}, {Poudel}, {Einasto},
  {Hein{\"a}m{\"a}ki}, {Lietzen}, {Tamm}, \& {Tempel}}]{2020A&A...639A..71K}
{Kuutma}, T., {Poudel}, A., {Einasto}, M., {et~al.} 2020, \aap, 639, A71

\bibitem[{{Lambas} {et~al.}(2003){Lambas}, {Tissera}, {Alonso}, \&
  {Coldwell}}]{2003MNRAS.346.1189L}
{Lambas}, D.~G., {Tissera}, P.~B., {Alonso}, M.~S., \& {Coldwell}, G. 2003,
  \mnras, 346, 1189

\bibitem[{{Ocvirk} {et~al.}(2006){Ocvirk}, {Pichon}, {Lan{\c{c}}on}, \&
  {Thi{\'e}baut}}]{2006MNRAS.365...74O}
{Ocvirk}, P., {Pichon}, C., {Lan{\c{c}}on}, A., \& {Thi{\'e}baut}, E. 2006,
  \mnras, 365, 74

\bibitem[{{Pan} {et~al.}(2012){Pan}, {Vogeley}, {Hoyle}, {Choi}, \&
  {Park}}]{2012MNRAS.421..926P}
{Pan}, D.~C., {Vogeley}, M.~S., {Hoyle}, F., {Choi}, Y.-Y., \& {Park}, C. 2012,
  \mnras, 421, 926

\bibitem[{{Park} {et~al.}(2023){Park}, {Chun}, {Shin}, {Jeong}, {Lee}, {Pak},
  {Smith}, \& {Kim}}]{2023ApJ...954...98P}
{Park}, S.-M., {Chun}, K., {Shin}, J., {et~al.} 2023, \apj, 954, 98

\bibitem[{{Pasquali} {et~al.}(2019){Pasquali}, {Smith}, {Gallazzi}, {De Lucia},
  {Zibetti}, {Hirschmann}, \& {Yi}}]{2019MNRAS.484.1702P}
{Pasquali}, A., {Smith}, R., {Gallazzi}, A., {et~al.} 2019, \mnras, 484, 1702

\bibitem[{{Patton} {et~al.}(2011){Patton}, {Ellison}, {Simard}, {McConnachie},
  \& {Mendel}}]{2011MNRAS.412..591P}
{Patton}, D.~R., {Ellison}, S.~L., {Simard}, L., {McConnachie}, A.~W., \&
  {Mendel}, J.~T. 2011, \mnras, 412, 591

\bibitem[{{Patton} {et~al.}(2013){Patton}, {Torrey}, {Ellison}, {Mendel}, \&
  {Scudder}}]{2013MNRAS.433L..59P}
{Patton}, D.~R., {Torrey}, P., {Ellison}, S.~L., {Mendel}, J.~T., \& {Scudder},
  J.~M. 2013, \mnras, 433, L59

\bibitem[{{Peng} {et~al.}(2010){Peng}, {Lilly}, {Kova{\v{c}}}, {Bolzonella},
  {Pozzetti}, {Renzini}, {Zamorani}, {Ilbert}, {Knobel}, {Iovino}, {Maier},
  {Cucciati}, {Tasca}, {Carollo}, {Silverman}, {Kampczyk}, {de Ravel},
  {Sanders}, {Scoville}, {Contini}, {Mainieri}, {Scodeggio}, {Kneib}, {Le
  F{\`e}vre}, {Bardelli}, {Bongiorno}, {Caputi}, {Coppa}, {de la Torre},
  {Franzetti}, {Garilli}, {Lamareille}, {Le Borgne}, {Le Brun}, {Mignoli},
  {Perez Montero}, {Pello}, {Ricciardelli}, {Tanaka}, {Tresse}, {Vergani},
  {Welikala}, {Zucca}, {Oesch}, {Abbas}, {Barnes}, {Bordoloi}, {Bottini},
  {Cappi}, {Cassata}, {Cimatti}, {Fumana}, {Hasinger}, {Koekemoer},
  {Leauthaud}, {Maccagni}, {Marinoni}, {McCracken}, {Memeo}, {Meneux}, {Nair},
  {Porciani}, {Presotto}, \& {Scaramella}}]{2010ApJ...721..193P}
{Peng}, Y.-j., {Lilly}, S.~J., {Kova{\v{c}}}, K., {et~al.} 2010, \apj, 721, 193

\bibitem[{P\'erez \& Granger(2007)}]{PER-GRA:2007}
P\'erez, F. \& Granger, B.~E. 2007, Computing in Science and Engineering, 9, 21

\bibitem[{{Rodr{\'\i}guez-Medrano} {et~al.}(2023){Rodr{\'\i}guez-Medrano},
  {Paz}, {Stasyszyn}, {Rodr{\'\i}guez}, {Ruiz}, \&
  {Merch{\'a}n}}]{2023MNRAS.521..916R}
{Rodr{\'\i}guez-Medrano}, A.~M., {Paz}, D.~J., {Stasyszyn}, F.~A., {et~al.}
  2023, \mnras, 521, 916

\bibitem[{{Rodr{\'\i}guez-Medrano} {et~al.}(2024){Rodr{\'\i}guez-Medrano},
  {Springel}, {Stasyszyn}, \& {Paz}}]{2024MNRAS.tmp..184R}
{Rodr{\'\i}guez-Medrano}, A.~M., {Springel}, V., {Stasyszyn}, F.~A., \& {Paz},
  D.~J. 2024, \mnras

\bibitem[{{Rojas} {et~al.}(2004){Rojas}, {Vogeley}, {Hoyle}, \&
  {Brinkmann}}]{2004ApJ...617...50R}
{Rojas}, R.~R., {Vogeley}, M.~S., {Hoyle}, F., \& {Brinkmann}, J. 2004, \apj,
  617, 50

\bibitem[{{Rojas} {et~al.}(2005){Rojas}, {Vogeley}, {Hoyle}, \&
  {Brinkmann}}]{2005ApJ...624..571R}
{Rojas}, R.~R., {Vogeley}, M.~S., {Hoyle}, F., \& {Brinkmann}, J. 2005, \apj,
  624, 571

\bibitem[{{Rosas-Guevara} {et~al.}(2022){Rosas-Guevara}, {Tissera}, {Lagos},
  {Paillas}, \& {Padilla}}]{2022MNRAS.517..712R}
{Rosas-Guevara}, Y., {Tissera}, P., {Lagos}, C. d.~P., {Paillas}, E., \&
  {Padilla}, N. 2022, \mnras, 517, 712

\bibitem[{{Scudder} {et~al.}(2012){Scudder}, {Ellison}, {Torrey}, {Patton}, \&
  {Mendel}}]{2012MNRAS.426..549S}
{Scudder}, J.~M., {Ellison}, S.~L., {Torrey}, P., {Patton}, D.~R., \& {Mendel},
  J.~T. 2012, \mnras, 426, 549

\bibitem[{{Taverna} {et~al.}(2023){Taverna}, {Salerno}, {Daza-Perilla},
  {D{\'\i}az-Gim{\'e}nez}, {Zandivarez}, {Mart{\'\i}nez}, \&
  {Ruiz}}]{2023MNRAS.520.6367T}
{Taverna}, A., {Salerno}, J.~M., {Daza-Perilla}, I.~V., {et~al.} 2023, \mnras,
  520, 6367

\bibitem[{{Tempel} {et~al.}(2017){Tempel}, {Tuvikene}, {Kipper}, \&
  {Libeskind}}]{2017A&A...602A.100T}
{Tempel}, E., {Tuvikene}, T., {Kipper}, R., \& {Libeskind}, N.~I. 2017, \aap,
  602, A100

\bibitem[{{V{\'a}squez-Bustos} {et~al.}(2023){V{\'a}squez-Bustos},
  {Argudo-Fernandez}, {Grajales-Medina}, {Duarte Puertas}, \&
  {Verley}}]{2023A&A...670A..63V}
{V{\'a}squez-Bustos}, P., {Argudo-Fernandez}, M., {Grajales-Medina}, D.,
  {Duarte Puertas}, S., \& {Verley}, S. 2023, \aap, 670, A63

\bibitem[{{Vazdekis} {et~al.}(2016){Vazdekis}, {Koleva}, {Ricciardelli},
  {R{\"o}ck}, \& {Falc{\'o}n-Barroso}}]{2016MNRAS.463.3409V}
{Vazdekis}, A., {Koleva}, M., {Ricciardelli}, E., {R{\"o}ck}, B., \&
  {Falc{\'o}n-Barroso}, J. 2016, \mnras, 463, 3409

\bibitem[{Virtanen {et~al.}(2020)Virtanen, Gommers, Oliphant, Haberland, Reddy,
  Cournapeau, Burovski, Peterson, Weckesser, Bright, {van der Walt}, Brett,
  Wilson, Millman, Mayorov, Nelson, Jones, Kern, Larson, Carey, Polat, Feng,
  Moore, {VanderPlas}, Laxalde, Perktold, Cimrman, Henriksen, Quintero, Harris,
  Archibald, Ribeiro, Pedregosa, {van Mulbregt}, \& {SciPy 1.0
  Contributors}}]{2020SciPy-NMeth}
Virtanen, P., Gommers, R., Oliphant, T.~E., {et~al.} 2020, Nature Methods, 17,
  261

\bibitem[{{Wetzel} {et~al.}(2013){Wetzel}, {Tinker}, {Conroy}, \& {van den
  Bosch}}]{2013MNRAS.432..336W}
{Wetzel}, A.~R., {Tinker}, J.~L., {Conroy}, C., \& {van den Bosch}, F.~C. 2013,
  \mnras, 432, 336

\end{thebibliography}

\end{document}